\documentclass[fleqn,usenatbib]{mnras}
\usepackage{newtxtext,newtxmath}
\usepackage[T1]{fontenc}

\DeclareRobustCommand{\VAN}[3]{#2}
\let\VANthebibliography\thebibliography
\def\thebibliography{\DeclareRobustCommand{\VAN}[3]{##3}\VANthebibliography}

\usepackage[utf8]{inputenc}
\usepackage{mathtools}
\usepackage{nameref}
\usepackage{wasysym}
\usepackage{makecell}
\usepackage{hyperref}
\usepackage[capitalise]{cleveref}
\usepackage{booktabs}
\usepackage{float}
\usepackage{graphicx}

\newcommand{\rev}[1]{#1}

\graphicspath{{img/}}

\title{Synthetic Evolution Tracks of Giant Planets}

\author[Simon Müller and Ravit Helled]{
    Simon Müller$^{1}$\thanks{E-mail: simon.mueller7@uzh.ch}
    and Ravit Helled$^{1}$
    \\
    $^{1}$Center for Theoretical Astrophysics and Cosmology \\
          Institute for Computational Science, University of Zürich \\
          Winterthurerstrasse 190, 8057 Zürich, Switzerland
}

\date{Accepted 2021 July 30. Received 2021 July 29; in original form 2021 June 4}
\pubyear{2021}

\begin{document}
\label{firstpage}
\pagerange{\pageref{firstpage}--\pageref{lastpage}}
\maketitle

\begin{abstract}
Giant planet evolution models play a crucial role in interpreting observations and constraining formation pathways. However, the simulations can be slow or prohibitively difficult. 
To address this issue, we calculate a large suite of giant planet evolution models using a state-of-the-art planetary evolution code. Using these data, we create the python program \texttt{planetsynth} that generates synthetic cooling tracks by interpolation. 
Given the planetary mass, bulk \& atmospheric metallicity, and incident stellar irradiation, the program calculates how the planetary radius, luminosity, effective temperature, and surface gravity evolve with time. 
We demonstrate the capabilities of our models by inferring time-dependent mass-radius diagrams, estimating  the metallicities from mass-radius measurements, and by showing how atmospheric measurements can further constrain the planetary bulk composition. We also estimate the mass and metallicity of the young giant planet 51 Eri b from its observed luminosity.
Synthetic evolution tracks have many applications, and we suggest that they are valuable for both theoretical and observational investigations into the nature of giant planets.
\end{abstract}


\section{Introduction}\label{sec:introduction}

Studying the internal structure and bulk compositions of gas giants is crucial for our understanding of giant planet formation and evolution. However, because giant planets are mostly composed of compressible hydrogen and helium, they contract and their observable properties and interiors change as the planets evolve \citep{Hubbard1977}. Accounting for the time-dependent nature properly requires comprehensive thermal evolution models. In order to better characterize giant planets, their the basic physical parameters such as their mass, bulk \& atmospheric composition, and stellar irradiation must be known \citep{Burrows2007,Thorngren2016}. While observations provide valuable information -such as occurrence rates, orbital distances, masses and radii - combining the measurements with theory is critical for the data interpretation and for taking full advantage of the measurements.

One key objective of giant planet science is to determine the planetary bulk composition from mass-radius measurements \citep{Fortney2007,Miller2011,Thorngren2016,Teske2019}. This is important because the predicted composition depends on how planets form \citep{Mousis2009,Johansen2017,Hasegawa2018,Ginzburg2020,Shibata2020} and can be used to narrow the range of possible formation pathways. An additional constraint is the planetary atmospheric composition that will be measured with high precision with the upcoming James Webb Space Telescope  (JWST; \citet{Gardner2006}) and ESA's Ariel mission \citep{Tinetti2018}
\rev{, which can be used to further constrain the planetary interior and to provide information on the link between bulk and atmospheric composition \citep{2019ApJ...874L..31T}}.
More precise radius and mass measurements, as well as an accurate determination of stellar ages are expected from the ESA Plato mission \citep{Rauer2014}. Improved measurements are also expected from ground-based facilities such as HARPS \citep{2003Msngr.114...20M}, NIRPS \citep{2017Msngr.169...21B} and ESPRESSO \citep{2021A&A...645A..96P}.

The upcoming data are expected to significantly improve our understanding of gaseous planets. However, the interpretation of the observations crucially depends on predictions from evolution models, which come with their own uncertainties:  Previous studies have shown that model parameters such as equations of state and atmospheric models \citep{Burrows2007,Baraffe2008,Vazan2013,Thorngren2016,Poser2019} influence the model predictions to a degree that is important as measurements become more accurate. Therefore, unyielding the full potential of these new observations requires careful modeling of the evolution of giant planets \citep{2020ApJ...903..147M}.

Another increasingly important application for evolution models is the mass determination of directly imaged exoplanets by, for example, the Hubble Space Telescope or SPHERE \citep{2019A&A...631A.155B}. In the near future, JWST is expected to provide many more detections. Direct imaging is unique in the sense that it measures the planetary luminosity, which then requires evolution models to infer the planet's mass \citep[e.g.,][]{2003A&A...402..701B,Marley2007,2017A&A...608A..72M}. For giant exoplanets, inferring the mass from luminosity measurements is commonly done by using pre-computed tables of planetary isochrones \citep{2003A&A...402..701B}. The disadvantage of this method is that it over-simplifies the planetary evolution suggesting that it is mostly the planetary mass that controls the luminosity while in fact there are several other important parameters that affect the planetary evolution \citep{2020ApJ...903..147M}.

Considering the wealth of new observations expected in the upcoming years, access to comprehensive evolution models is essential. With that in mind, we have calculated an extensive set of giant planet evolution models using a modified version of Modules for Experiments in Stellar Astrophysics (MESA; \citet{Paxton2011,Paxton2013,Paxton2015,Paxton2018,Paxton2019}). These evolution models were then used to build \texttt{planetsynth} (\url{https://github.com/tiny-hippo/planetsynth}), a publicly available python program that generates synthetic evolution tracks that account for the planetary mass, bulk \& atmospheric metallicity and incident stellar irradiation. Given these parameters, \texttt{planetsynth} calculates how the planet's radius, luminosity, effective temperature and surface gravity evolve with time.

This paper is organised as follows: in \S \ref{sec:thermal_evolution}, we present the planetary evolution calculations and describe the most important model assumptions and their influences. Then, we present the parameter space covered by our suite of evolution models in \S \ref{sec:suite}. We explain how the synthetic cooling tracks are generated in \S \ref{sec:cooling_tracks}, and validate our approach in \S \ref{sec:validation}. We then propose key possible implications for the synthetic cooling tracks, and demonstrate their use on some real-world examples in \S \ref{sec:applications}. Key assumptions and limitations of our models as well as future work are discussed in \S \ref{sec:discussion}. A summary of our work is given in \S \ref{sec:conclusions}.

\section{Thermal Evolution of Giant Planets}\label{sec:thermal_evolution}

Giant planets start their lives hot and inflated \citep{Hubbard1977,Cumming2018,Valletta2020}. Since their interiors are compressible, they contract as they cool down. To calculate the cooling of giant planets, we use a modified version of the stellar evolution code MESA. It solves the 1D hydrostatic equilibrium evolution equations \citep[e.g.,][]{Kippenhahn2012} and calculates the evolution of the interior structure and of observable quantities such as the planetary radius, luminosity, and effective temperature.

The four parameters we use to create the planetary models are mass $M$, bulk metallicity $Z$, envelope metallicity $Z_{env}$ and incident stellar irradiation $F_*$ (see \S \ref{sec:appendix_model_creation} for details on how the models were created with MESA). The bulk composition is defined by the hydrogen, helium, and heavy-element mass fractions. We assume a proto-solar hydrogen-helium ratio ($X_{proto} = 0.705$, $Y_{proto} = 0.275$), which defines $Z$ as the free parameter that sets the planetary bulk composition. 

For planets smaller than $5 M_J$ and with $Z \neq Z_{env}$, we further assume a core-envelope structure, i.e., the heavy-elements are mostly contained in a compact core in the deep interior. It should be noted, however, that formation models clearly show that the deep interiors of giant planets consist of composition gradients or more extended and dilute cores \citep{Lozovsky2017,Helled2017,Valletta2020}. Nevertheless, this region is typically relatively small. Furthermore, evolution models suggest that composition gradients are quickly destroyed by large-scale convection \citep{Mueller2020} unless the internal temperatures are relatively low \citep{Vazan2018}.

For planets beyond $5 M_J$ or with $Z = Z_{env}$ a homogeneous composition is used. The heavy-element distribution directly affects the radius and luminosity of the planet \citep{Baraffe2008,Vazan2013}. The effect is, however, small compared to the choice of the equation of state (hereafter EoS) or the material that is used to represent the heavy elements \citep{2020ApJ...903..147M}, in particular for massive giant planets.

In order to self-consistently account for the heavy elements in the evolution calculations, we made substantial changes to the EoS of MESA as presented in \citet{Mueller2020}. The modifications to the EoS make it suitable to model giant planets consisting of arbitrary mixtures of hydrogen, helium and a heavy element. For this study, we have added a 50-50 water-rock mixture, which we use in all our models to represent the heavy elements. \rev{More details on the EoS are given in \S \ref{sec:appendix_equation_of_state}}. Other than the EoS, both stellar irradiation and atmospheric metallicities both have a large influence on the evolution \citep[e.g.,][]{Burrows2007,2020ApJ...903..147M}; we include both of these as model parameters (\rev{see \S \ref{sec:appendix_atmosphere} for details}).

Our calculations assume that giant planets form by core accretion \citep{Mizuno1980,Pollack1996} in the \textit{hot-start} framework. In hot-start models the planets cool adiabatically from an initially large, but essentially arbitrary radius \citep[e.g.,][]{Marley2007}. As described above, we model the planets with either homogeneous compositions or core-envelope structures. In that case, it is reasonable to assume that the planets cool adiabatically for most of their lives, and different initial conditions converge quickly. It must be noted, however, that current studies of Jupiter and Saturn suggest that their interiors are - at least in part - non-adiabatic \citep{Debras2019,2021arXiv210413385M}, and we hope to address this in future research. We discuss the implications of using adiabatic hot-start models in more detail in \S \ref{sec:discussion} and \S \ref{sec:appendix_initial_conditions}.

\subsection{Suite of Giant Planet Evolution Models}\label{sec:suite}
Giant exoplanets have large ranges of masses and compositions and are irradiated by their host stars with fluxes that can range many order of magnitudes \citep{Thorngren2016,Teske2019}, and the suite of planetary evolution models presented here needs to account for this diversity. The evolution parameters are sampled from a non-uniform four-dimensional grid spanned by the planetary mass $M$, bulk heavy-element fraction $Z$, atmospheric metallicity $Z_{env}$ and incident stellar flux $F_*$. The planetary masses we consider range from $0.1 - 30 M_J$. For planets more massive than about a Jupiter mass, the planetary radius is only changing slightly as a function of mass. Therefore, we sample the masses more coarsely beyond $1 M_J$. \rev{Note that the mass limit for the ignition of deuterium is generally believed to be at about $13 M_J$ \citep[e.g.,][]{2007IAUTA..26..183B}. However, since this would mostly affect very young and massive planets \citep{2011ApJ...727...57S} for which uncertainties in the formation luminosity are high \citep[e.g.,][]{2016PASP..128j2001B}, we do not include deuterium burning in this work.}

Depending on the mass, the grid covers different ranges of $Z$ because i) the bulk metallicity tends to decrease with planetary mass and ii) there are limitations in the EoS. Nonetheless, this provides coverage for most of the inferred compositions of observed giant exoplanets population \citep{Thorngren2016}.

\begin{table}
    \centering
    \caption[Planetary Parameters]{Masses $M$, bulk metallicities $Z$, atmospheric metallicities $Z_{env}$ and incident stellar fluxes $F_*$ covered by the suite of models. In total, the simulation grid is composed of about 45,000 planets.}
    \begin{tabular}{cccc}
    \toprule
        $M$ [M$_J$] & $Z$ & $Z_{env}$ & $F_*$ [erg/s/cm$^{-2}$]  \\ \midrule
        $0.1 - 1$   &$\le 0.80$ &$\le 0.10$ &$10 - 10^{9}$  \\
        $1 - 3$     &$\le 0.50$ &$\le 0.10$ &$10 - 10^{9}$  \\
        $3 - 5$     &$\le 0.20$ &$\le 0.10$ &$10 - 10^{9}$  \\
        $5 - 30$    &$\le 0.04$ &$\le 0.04$ &$10 - 10^{9}$ \\ \bottomrule
    \end{tabular}

    \label{table:planet_params}
\end{table}

For the stellar flux, we consider the range from $10 - 10^{9}$ erg s$^{-1}$ cm$^{-2}$. This includes cool to warm Jupiters, but excludes extremely irradiated hot Jupiters, since their inflation mechanism remains unknown \citep[e.g.,][]{Fortney2010,Weiss2013,Baraffe2014}. \rev{The conventional cut-off irradiation for hot Jupiters  is estimated to be $2 \times 10^{8}$ erg s$^{-1}$ cm$^{-2}$ \citep{Miller2011}. At the maximum flux that we consider, observations suggest that exoplanets can inflate to $\simeq 1.5 R_J$ \citep{2018AJ....155..214T}.  Therefore our evolution tracks with stellar fluxes beyond $2 \times 10^{8}$ erg s$^{-1}$ cm$^{-2}$ should be treated with caution}. We do not include photo-evaporation, which is expected to  become important for the lowest masses and highest stellar fluxes we consider \citep[e.g.,][]{2014ApJ...795...65J}. All the parameters covered by our suite of models are summarised in Table \ref{table:planet_params}. In total, our simulation grid consists of about 45,000 planets.

The thermal evolution of the planets in the sample is calculated as described in \S \ref{sec:thermal_evolution} for $10^{10}$ years. The relevant outputs are the planetary radius $R$, luminosity $L$, effective temperature $T_{e}$ and surface gravity $g$ as they evolve with time.
\rev{The luminosity corresponds to the heat from the interior without any re-radiated stellar irradiation; the irradiation heats the outer layers of the planet and increases the luminosity at the surface (see \S \ref{sec:discussion} and \S \ref{sec:appendix_atmosphere} for details).} 
As discussed above, at early times the evolution could be significantly affected by the initial condition, and hence we exclude data for times below $10^7$ years.

For a few parameters the evolution calculations crashed before reaching the end, because some temperatures and densities inside the planet are outside of the EoS boundaries. This occurs mostly for planets at the edges of our simulation grid towards the end of their cooling. However, after about 1 Gyr the planetary radius and luminosity change only very slowly with time. To be on the safe side, we discarded models that crashed before 2 Gyr. If the calculations fail between 2-10 Gyr, we extrapolate the subsequent luminosity and radius evolution with a function of the form $f(t) = a \log(t) + b$, where $a$ and $b$ are parameters that are individually fitted to the planetary parameters using its evolution history. The effective temperature and surface gravity are then calculated from the extrapolated luminosity and radius. This approach leads to errors typically smaller than 0.1\% in all output quantities, which is negligible compared to both observational and theoretical uncertainties.

To summarise: we simulate a suite of planetary models covering a large range of masses, bulk \& envelope metallicities and incident stellar fluxes. The calculations determined the radius, luminosity, effective temperature and gravitational acceleration at the surface $\left(R(t),\, L(t),\, T_{e}(t),\, g(t)\right)$ as a function of the planetary age $t$.

\subsection{Generating Synthetic Cooling Tracks}\label{sec:cooling_tracks}

The suite of giant planet evolution models enables the generation of synthetic cooling tracks for a given set of planetary parameters (within the appropriate limits, see Table \ref{table:planet_params}). We have thoroughly tested several approaches, such as directly interpolating on the irregular grid or using machine-learning regression with XGBoost \citep{Chen:2016:XST:2939672.2939785} and neural networks. The former is too slow when making a large number of predictions. The latter are fast and generally yield accurate outputs. However, ensuring that the predictions were physically consistent is difficult. 

We have therefore settled on the following approach:

\begin{enumerate}
    \item In order to generate data of equal length, a cubic spline interpolation is used to calculate the outputs (radius, luminosity and effective temperature) for each planet at discrete times between $10^7 - 10^{10}$ years. The interpolated data from all planets is then combined into a single data set.
    \item While it would have been possible to directly interpolate this data to generate new cooling tracks, it is quite slow for a four dimensional parameter space on an irregular grid. Therefore, we first map our data onto a four-dimensional regular, unevenly spaced grid using piece-wise linear interpolation. For the regions in parameter space where we do not have data (e.g., high-mass and metallicity planets), we use nearest neighbour interpolation. While this may be a decent estimate for parameters that are just outside the range covered by our suite of models, the errors will quickly grow large. Therefore, these data points should never be used to calculate predictions, but rather they were a convenient way to create a regular grid. \rev{The default behaviour when attempting to generate synthetic cooling tracks for parameters not covered by our model suite is to warn the user and return NaNs.}
    \item The synthetic cooling tracks are then generated by linearly interpolating on the regular grid. For each set of input parameters ($M,\, Z,\, Z_{env},\, \log F_*$), the interpolation yields ($R \, [R_J], \log L \, [L_\odot]$) at discrete times between $\log t = 7 - 10$ [yrs]. In order to provide self-consistent results, we calculate the effective temperature from the generated radius and luminosity, assuming a black body: $T_{e}^4 = L / 4 \pi \sigma R^2$ where $\sigma$ is the Stefan-Boltzmann constant. The gravitational acceleration is calculated from the generated radius as $\log g = \log\left(\textrm{G} M / R^2\right)$ [cm/s$^2]$.
    \item If the prediction is required at specific time $\log t$ [yr], a cubic spline interpolation is calculated from the synthetic cooling track.
\end{enumerate}

Our approach is similar to that of the MESA Isochrones and Stellar Tracks (MIST) \citep{2016ApJ...823..102C,2016ApJS..222....8D} project, which uses a grid of stellar evolutionary tracks and isochrones to generate cooling tracks for stars. The last two steps described above were implemented as the open-source python module called \texttt{planetsynth}, with the goal of it being both easy to use and update. Another advantage is that it is very fast: a million synthetic cooling tracks can be generated in a just a few seconds (for vectorized inputs).

\subsection*{planetsynth: A Fast and and Easy-to-Use Python Module for Generating Synthetic Cooling Tracks}

The main class of the module is \texttt{PlanetSynth}, which contains the two most important methods \texttt{synthesize} and \texttt{predict}.

\begin{itemize}
    \item \texttt{synthesize}: takes the planetary parameters $\left(M,\, Z,\, Z_{env},\, \log F_*\right)$ as arguments and calculates the cooling track at discrete times between $10^7 - 10^{10}$ years.
    \item \texttt{predict}: uses single or several points in time to calculate $\left(R \, , \log L , \, T_{e}, \, \log g \right)$ at those specific times for a given set of planetary parameters.
\end{itemize}

The module is available at \url{https://github.com/tiny-hippo/planetsynth} and includes examples in some detail on how to use it.

\section{Results}\label{sec:results}

In this section we first show that the synthetic cooling tracks yield accurate results when compared to the direct calculations with MESA. Then, we demonstrate the capabilities of our code by calculating time-dependent mass-radius diagrams, inferring metallicities from mass-radius measurements, and inferring the mass and metallicity of a directly imaged planet.

\begin{figure}
    \centering
    \includegraphics[width=\columnwidth]{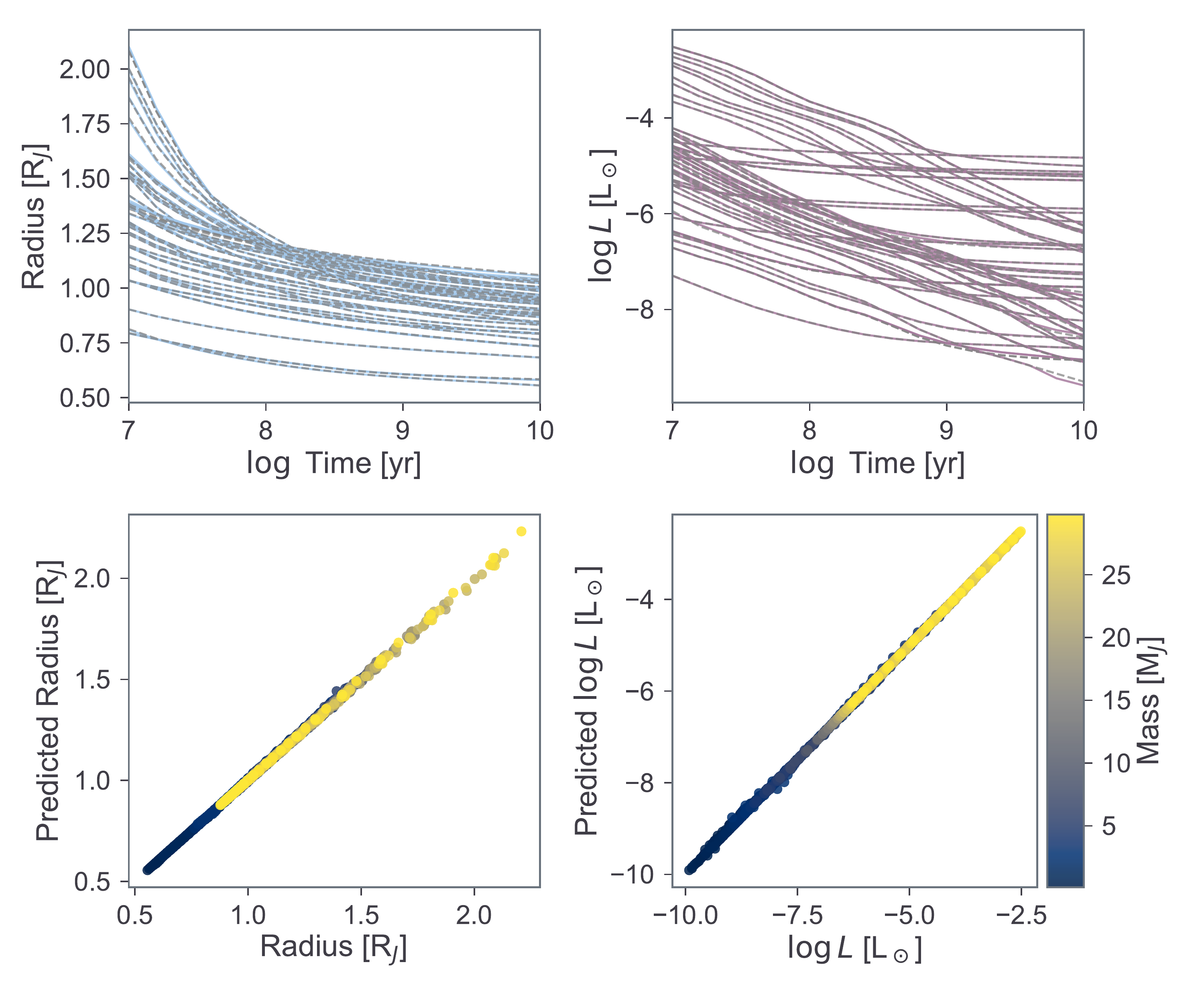}
    \caption{\textbf{Top:} Radius (left) and luminosity (right) evolution for 50 randomly chosen planets from the validation sample. The solid lines are the synthetic cooling tracks, while dashed lines show the predictions from the evolution calculations. \textbf{Bottom:} Calculated vs.~predicted radius (left), luminosity (centre) and effective temperature (right) for the full validation sample (coloured by mass).}
    \label{fig:R_logL_grid}
\end{figure}

\subsection{Validation}\label{sec:validation}

In order to validate our approach, we create a sample of 200 planets (see \S \ref{sec:appendix_performance_testing}). The planetary parameters are generated with the Latin Hypercube sampling method \citep{McKay1979}, which is a statistical method commonly used in high-dimensional numerical experiments. Since the validation sample was not included when fitting the interpolants, it can be used as a non-biased sample to validate the performance of the synthetic models.

For each planet in the validation sample, we calculate the thermal evolution and then compare the output to the predictions from the synthetic cooling tracks. For a qualitative comparison, we plot the radius and luminosity evolution of 50 randomly chosen planets from the validation sample in the top panels of \cref{fig:R_logL_grid}. It can be seen that the synthetic cooling tracks agree extremely well with those from the evolution calculations. In the bottom panels of \cref{fig:R_logL_grid}, we plot the radii and luminosities from the evolution models against the synthesised predictions for the entire validation sample: The agreement is excellent, with only small deviations from the actual values. For a more detailed discussion of the model performance and the error distributions see \S \ref{sec:appendix_performance_testing}.

\subsection{Examples of Scientific Applications}\label{sec:applications}

\subsubsection{Time-Dependent Mass-Radius Diagrams}\label{sec:mass_radius}

Mass-Radius diagrams are often used to interpret the possible compositions of exoplanets, and to qualitatively compare exoplanets to our solar system planets. While the radii of planets made of refractory materials are not expected to change much with time, this is very different for gas-rich planets that contract significantly with time. The planetary age cannot be ignored when constructing mass-radius diagrams of giant planets since the radius depends on both mass and time.

\begin{figure}
    \centering
    \includegraphics[width=\columnwidth]{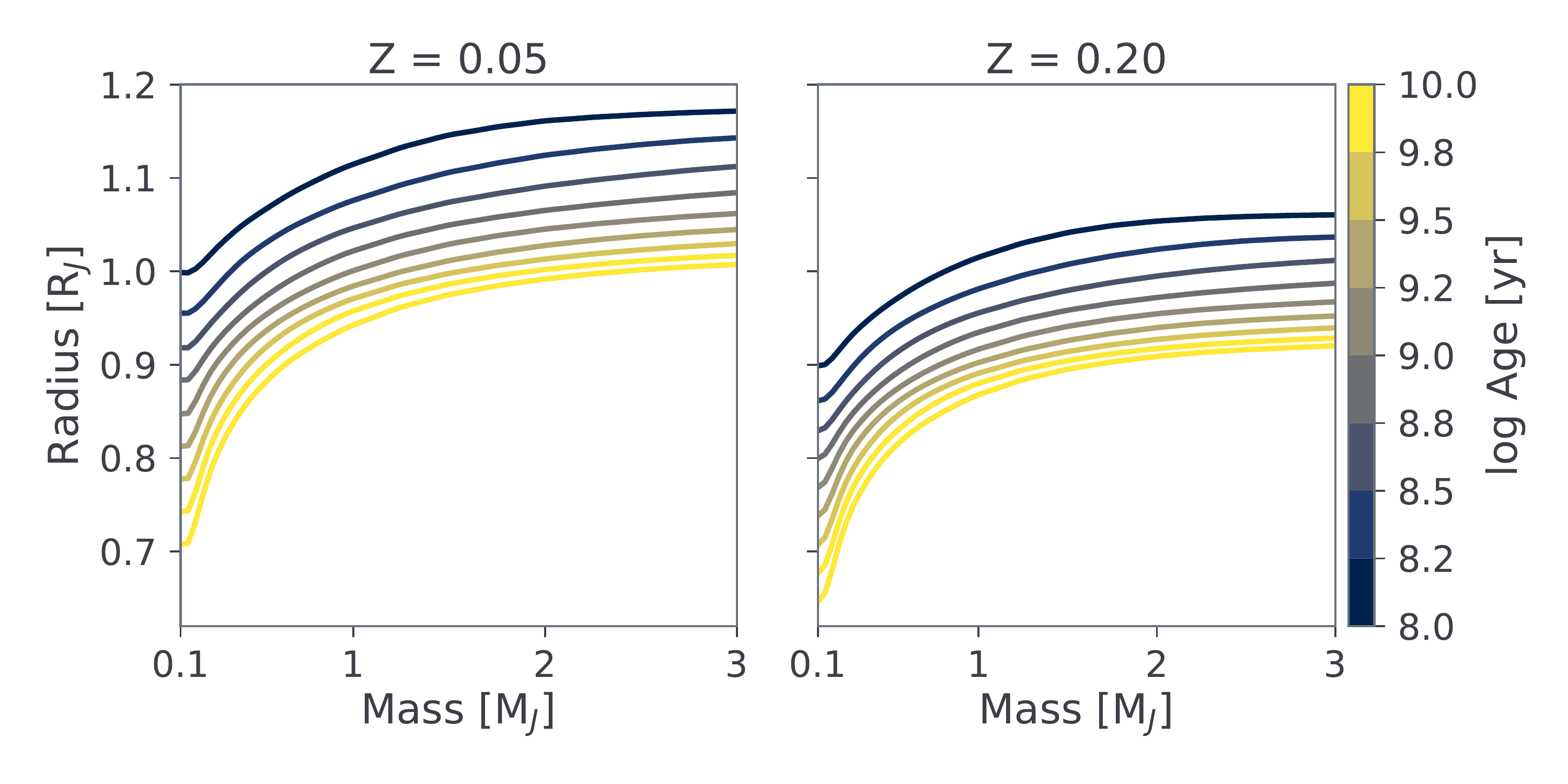}
    \caption{Time-dependent mass-radius diagrams calculated from synthetic cooling tracks using $Z = 0.05$ (left) and $Z = 0.20$ (right), stellar atmospheric metallicity and incident flux of $F_* = 10^5$ erg/s/cm$^2$. The colours correspond to isochrones from $10^8$ to $10^{10}$ years.}
    \label{fig:mass_radius_diagram}
\end{figure}

Using our synthetic evolution tracks it is simple to calculate mass-radius diagrams that account for the time dependency of the radius. We show two examples in \cref{fig:mass_radius_diagram} with $Z = 0.05$ (left) and $Z = 0.20$ (right). In both cases, the incident stellar irradiation was $F_* = 10^5$ erg/s/cm$^2$ and the atmospheric metallicity was assumed to be solar. Plotting the figures side-by-side makes it clear that planets with large differences in composition can have the same radius at a different time. This will be particularly important when interpreting measurements from Plato, since it will provide accurate stellar ages with an uncertainty of $\sim 10 \%$ \citep{Rauer2014}.

\subsubsection{Inferring the Bulk Metallicity based on Mass-Radius Measurements}\label{sec:inferred_metallicity}

Constraining the bulk compositions of giant exoplanets is critical for understanding giant planet formation models. While the bulk composition cannot be measured directly, it can be inferred from mass-radius measurements (from radial velocity and transit observations) with the help of evolution models \citep{Guillot2006,Miller2011}. Here, instead of using evolution models directly, we use the generated evolution tracks. We follow the approach described in \citet{Thorngren2016}: for a given mass, age, stellar flux, and atmospheric metallicity we search for the bulk metallicities that fit the observed radius. We demonstrate this procedure on two giant exoplanets: i) Kepler-16 b with $M = 0.33 \pm 0.02 M_J$, $R = 0.754 \pm 0.003 R_J$, $F_* = 4.19 \times 10^5$ erg/s/cm$^2$, with an age between $0.5 - 10$ Gyr \citep{2011Sci...333.1602D}, and ii) HAT-P-54b with $M = 0.76 \pm 0.03 M_J$ and $R = 0.94 \pm 0.03 R_J$, $F_* = 1.04 \times 10^{8}$ erg/s/cm$^2$ and an age of $1.8 - 8.2$ Gyr \citep{2015AJ....149..149B}.
 
\begin{figure}
    \centering
    \includegraphics[width=\columnwidth]{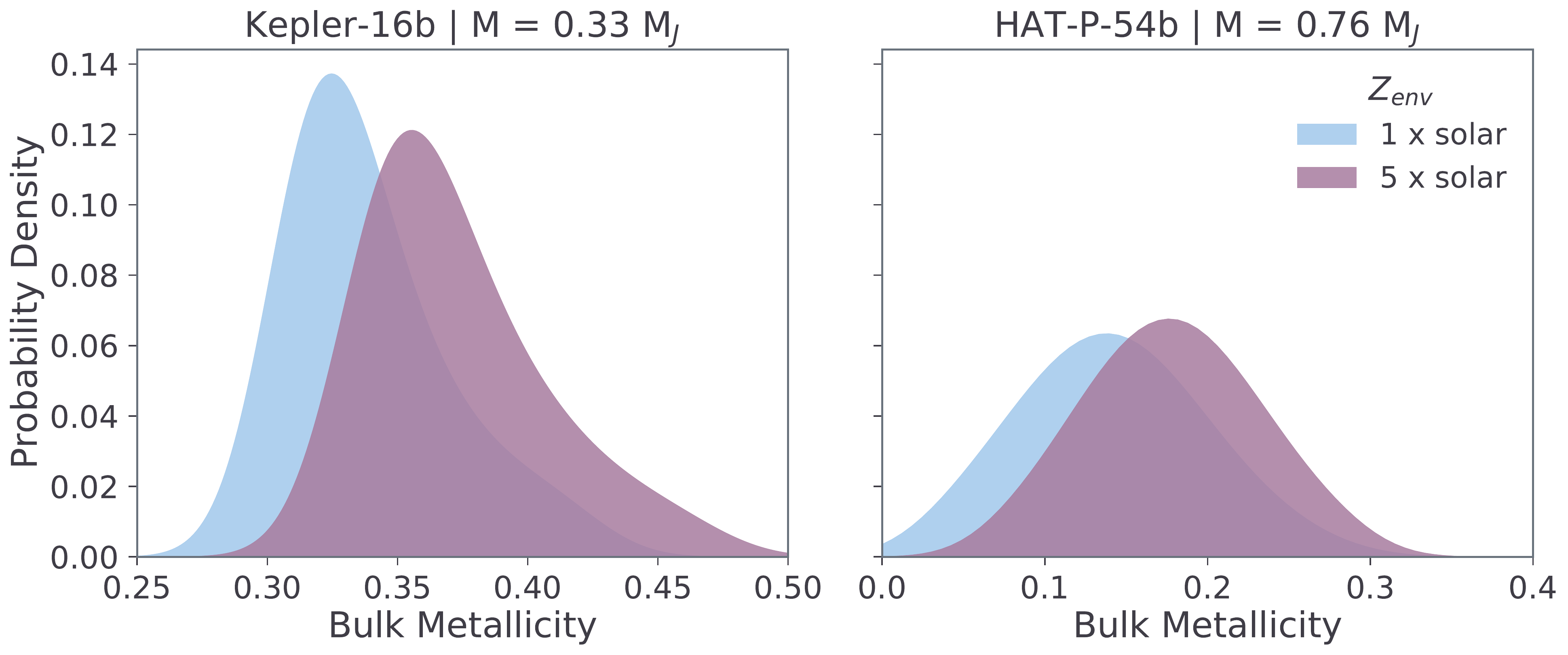}
    \caption{Inferred bulk metallicity assuming one (blue) and five (purple) times solar atmospheric metallicity of Kepler-16b and HAT-P-54b using a Gaussian KDEs constructed from 100,000 samples for each planet and atmospheric composition. The samples were calculated by generating synthetic cooling tracks that reproduce the observed radius.}
    \label{fig:inferred_metallicites}
\end{figure}

In order to infer the bulk metallicity, samples are drawn from the probability distributions for the measured planetary mass, radius and age. We assume normal distributions for the mass and radius, and a uniform distribution for the age. For each draw, we use a root-finding algorithm and our synthetic cooling tracks to find the metallicity that reproduces the observed radius. This is done twice for each planet with different atmospheric metallicities: one or five time solar. We repeat the procedure 100,000 times, which yields the estimates and their associated uncertainties of the heavy-element contents. The Gaussian kernel density estimates (KDEs) based on these samples are shown in \cref{fig:inferred_metallicites}. Note that we limit the x-axis on both plots to exclude low-probability compositions for better visual clarity.

For Kepler-16b the mean inferred metallicity is $Z = 0.34$ (SD = 0.03) and $Z = 0.37$ (SD = 0.03) for the solar and 5x solar atmospheric compositions, respectively. For HAT-P-54b, the values are $Z = 0.14$ (SD = 0.08) and $Z = 0.18$ (SD = 0.05). The mean and standard deviations are calculated directly from the samples rather than from the KDEs. While the results for both atmospheric models are within their respective uncertainties, there is a clear trend towards higher heavy-element contents when the atmosphere is enriched in heavy elements. This is because the atmosphere becomes more opaque as heavy elements are added, and therefore the planet is hotter at a given age and can be more metal-rich \citep{Burrows2007,2020ApJ...903..147M}. Note that our inferred metallicities for the solar atmosphere are significantly lower than previously reported estimates: $Z = 0.39^{+0.01}_{-0.02}$ for Kepler-16b and $Z = 0.20 \pm 0.04$ \rev{for HAT-P-54b} \citep{Thorngren2016}. This can mostly be attributed to the different hydrogen-helium EoSs \citep{2020ApJ...903..147M}.

\subsubsection{Further Constraining the Bulk Composition using Atmospheric Measurements}\label{sec:inferred_metallicity_atm}

\begin{figure}
    \centering
    \includegraphics[width=\columnwidth]{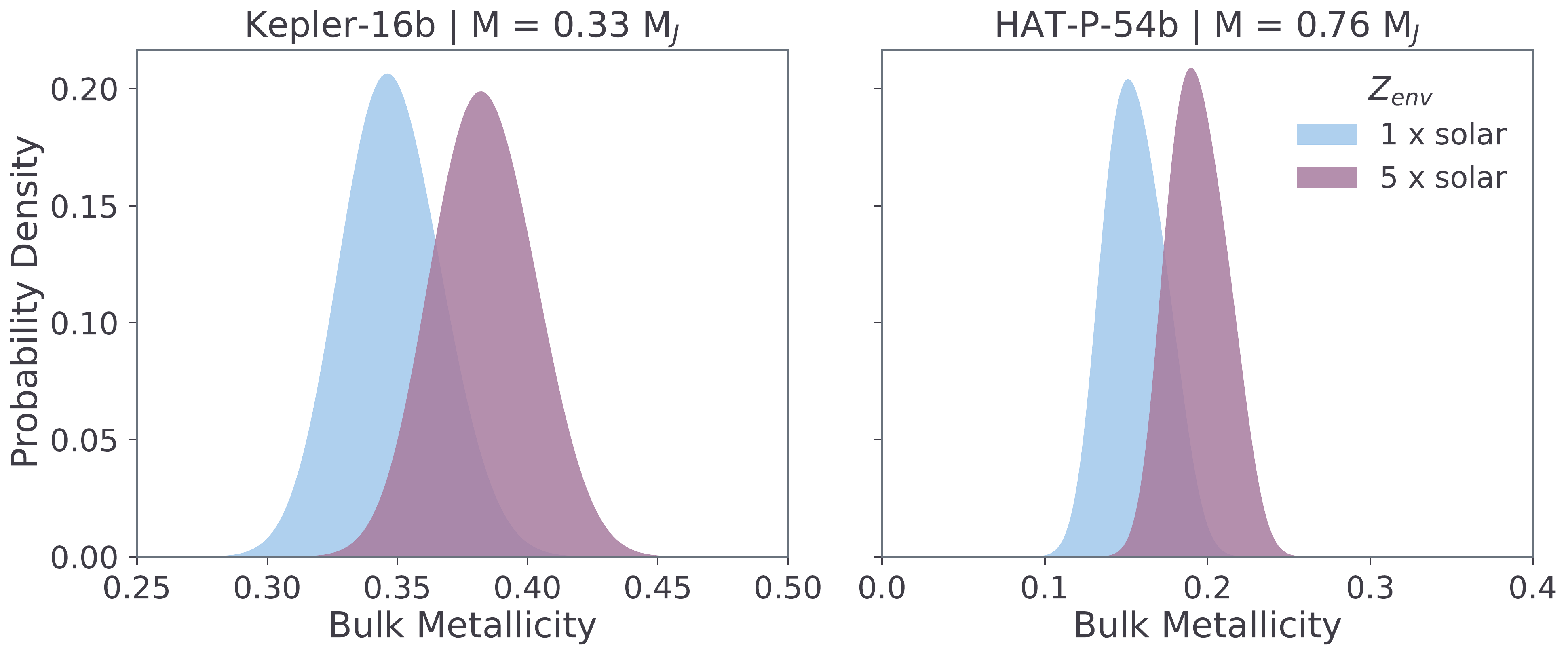}
    \caption{Same as \cref{fig:inferred_metallicites} but assuming $\Delta M = 0.005 M_J$ and $\Delta R = 0.005 M_J$ for the mass/radius uncertainties, and ages are constrained between $2 - 5$ Gyr.}
    \label{fig:inferred_metallicites_errors}
\end{figure}

Both Kepler-16b and HAP-P-54b have uncertainties in mass and/or radius of up to a few percent, while the age is rather poorly constrained. In this case, measuring an atmospheric metallicity would not help to constrain the bulk composition. However, it is interesting to investigate whether this remains true if measurements were to improve. Therefore, we repeat the above calculations, but with mass and radius uncertainties of $\Delta M = 0.005 M_J$ and $\Delta R = 0.005 M_J$, and ages constrained between $2 - 5$ Gyr. The resulting distributions are shown in \cref{fig:inferred_metallicites_errors}. Now, the inferred metallicities are $Z = 0.34, 0.38$ (SD = 0.02, 0.02) for Kepler-16b and $Z = 0.16, 0.19$ (SD = 0.02, 0.02) for HAT-P-54b, respectively.

For both planets the peaks in the probability density for the metallicity are now well-separated. This demonstrates that with improved observational uncertainties, measuring the atmospheric metallicity becomes a key ingredient in the interpretation of observations and can significantly improve the characterisation of giant exoplanets.

\subsubsection{Inferring the Mass and Bulk Metallicity from Luminosity Measurements}\label{sec:inferred_mass_metallicity}

In the examples so far, we have assumed that the planet's radius and mass are known. Alternatively, a giant planet can be observed by detecting its thermal emissions. In this case, the planet's luminosity (and atmospheric composition) are known, and the planetary mass can be inferred by comparing the observed luminosity with theoretical, mass-dependent evolutionary tracks \citep[e.g.,][]{2003A&A...402..701B}.

Here, we use synthetic cooling tracks to infer the mass and metallicity of the young giant planet 51 Eri b. Its observed luminosity is $\log L [L_\odot] = -5.8$ to $ -5.4$, its distance from its star is $\sim13$ au  and its mass was estimated to be $\sim 2 M_J$ (assuming a hot start) \citep{2015Sci...350...64M}. We follow a similar Monte Carlo approach as described in \S \ref{sec:inferred_mass_metallicity}. We draw luminosities from a normal distribution with a mean value of $\log L [L_\odot] = -5.6$ and a standard deviation of $\Delta \log L [L_\odot] = 0.2$. The age is inferred from a uniform distribution between $17 - 23$ Myr. For the atmospheric metallicity, we use the measured host star [Fe/H], which is roughly solar.

\begin{figure}
    \centering
    \includegraphics[width=\columnwidth]{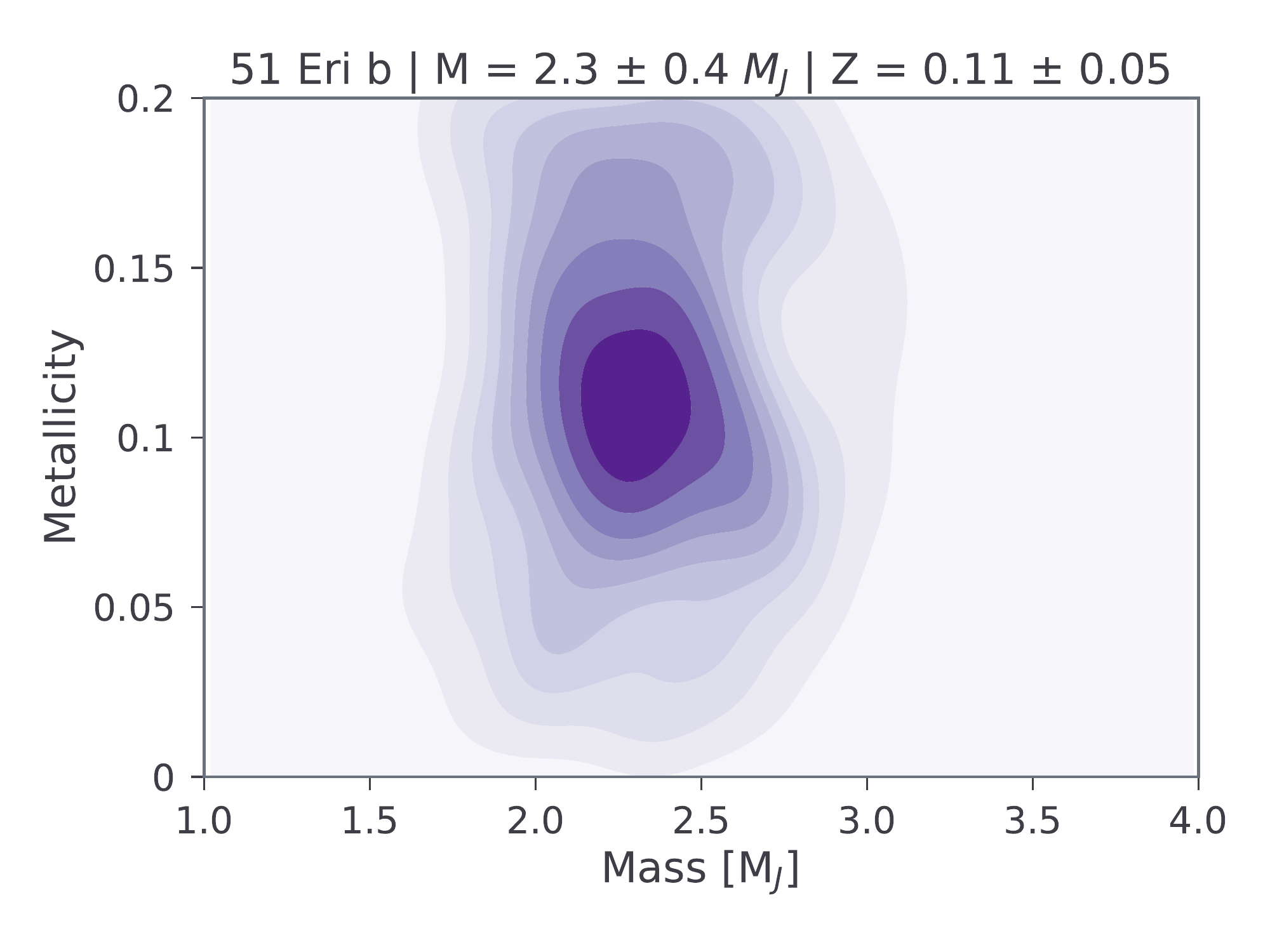}
    \caption{Inferred mass and bulk metallicity of 51 Eri b using a 2d Gaussian kernel density estimate based on 10,000 samples. \rev{The levels correspond to the probability of a point lying below a given contour within the interval [0.1, 0.9] in steps of 0.1.} The samples are calculated by generating synthetic cooling tracks that reproduce the inferred planetary luminosity.}
    \label{fig:51erib_kdeplot}
\end{figure}

\begin{figure}
    \centering
    \includegraphics[width=\columnwidth]{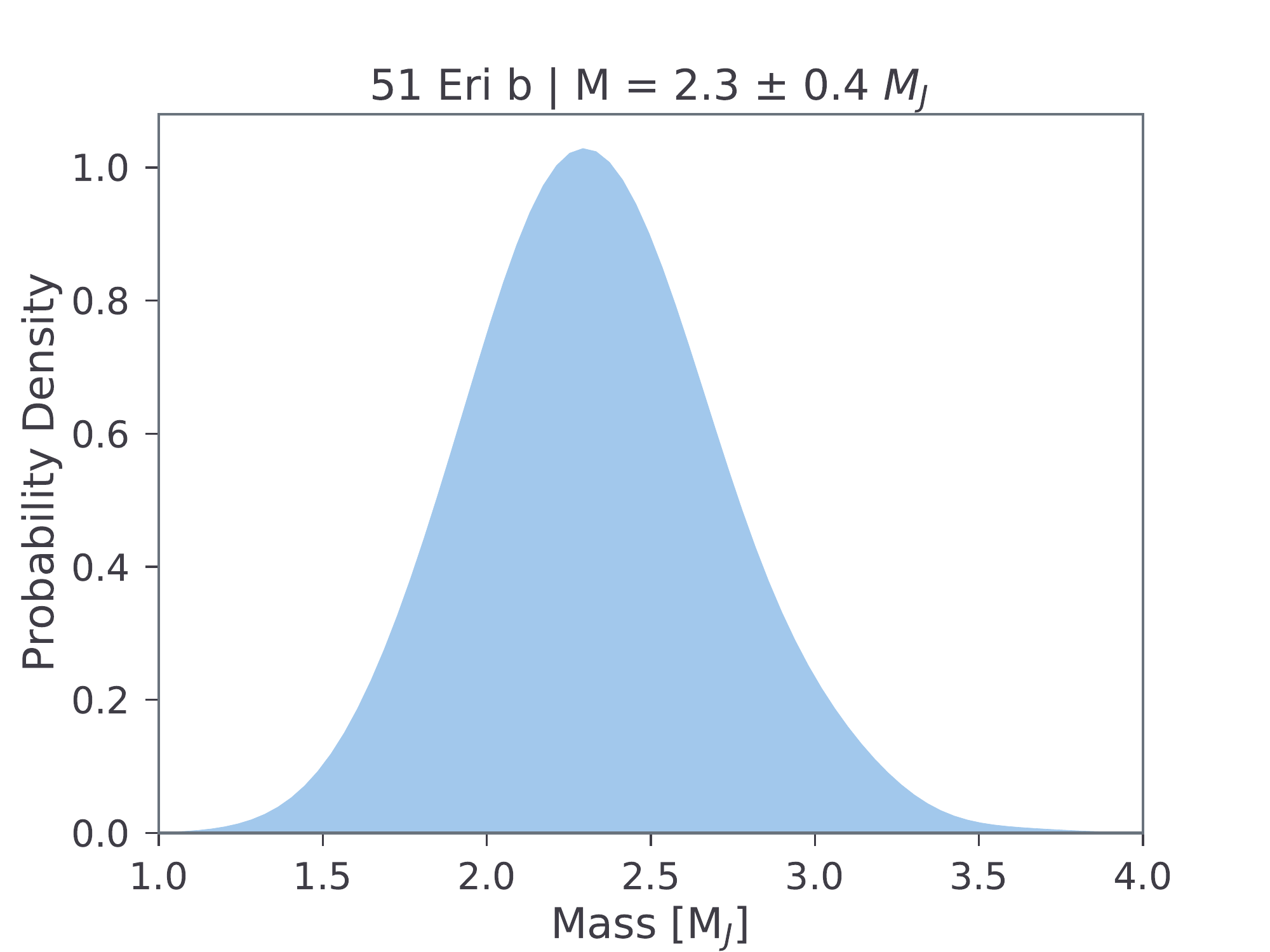}
    \caption{Inferred mass of 51 Eri b using a 1d Gaussian kernel density estimate based on 10,000 samples. The samples were calculated by generating synthetic cooling tracks that reproduce the inferred planetary luminosity.}
    \label{fig:51erib_histogram}
\end{figure}

For each sample, we use a multivariate optimization scheme to find the mass and bulk metallicity that closely matches the observed luminosity. We repeat this procedure 10,000 times and use a 2d Gaussian kernel density estimate to infer the probability distribution in the mass-metallicity plane. The resulting two-dimensional distribution is shown in \cref{fig:51erib_kdeplot}. In \cref{fig:51erib_histogram}, we show the 1d Gaussian kernel density estimate for mass of 51 Eri b.

The mass distribution is a slightly right-skewed Gaussian, with a mean value of $M = 2.3 \, M_J$ (SD = $0.4 \, M_J$) for the mass and $Z = 0.11$ (SD = $0.05)$ for the bulk metallicity. The mean mass estimate is higher than the $2 \, M_J$ reported in \citet{2015Sci...350...64M}, which can be explained by different model assumptions, such as equations of state and atmospheric models.

It is interesting to note that the most probable mass $M \simeq 2.3 M_J$ suggests a rather high bulk metallicity of $Z \simeq 0.11$, which corresponds to a heavy-element mass of roughly $M_Z \simeq 70 M_{\oplus}$. Large heavy-element masses are difficult to explain with current planet formation models since cores of gas giants are currently thought to be limited in size \citep[e.g.,][]{Bitsch2018}. Additional super-stellar metallicities could be explained by the accretion of planetesimals \citep{Mousis2009,Shibata2020} or pebbles \citep{Johansen2017}, or collisions between primordial cores \citep{Ginzburg2020}. However, it is still unclear whether these processes could explain highly-enriched gas giants. It is also important to re-state that our models are based on a hot-start formation scenario. Assuming a cold-start, luminosities at a given mass could be orders of magnitude lower (see \S \ref{sec:appendix_initial_conditions}), which would yield a mass estimate of $M \simeq 10 M_J$ \citep{2015Sci...350...64M}.

\section{Discussion}\label{sec:discussion}

While this study represents an advancement in evolution calculations, several aspects in the models can clearly be improved. To start with, future work could provide a higher resolution for the planetary parameters, which would further reduce the interpolation errors. \rev{For the massive planets, our models neglect deuterium burning, which could affect the luminosity and radius of young and massive planets  \citep{2011ApJ...727...57S}. We hope to include this in a future update of our evolution models. We also plan to include various  opacity and EoS tables since  they can influence the exoplanetary radius predictions \citep{Guillot1999}.}

Another aspect is the atmospheric model: while the atmospheric opacity accounts for heavy-element enrichment and the effects of stellar irradiation are included, we do not consider the effect of clouds or grains. Atmospheres that are cloudy or enriched in grains are more opaque, and trap the primordial heat more efficiently. This leads to a delayed cooling, which can significantly affect the evolution of a planet \citep{Vazan2013,Poser2019}, in particular at young ages. \rev{Another aspect that can be improved is the treatment of stellar irradiation: while the method we used in this study is in agreement with more sophisticated  treatments for planets that are not extremely irradiated \citep{2015ApJ...813..101V} it can clearly be improved. Future work should also include semi-grey atmospheric models \citep[e.g.,][]{Guillot2010,2014A&A...562A.133P} or full atmospheric models that include irradiation \citep[e.g.,][]{Fortney2007} as well as different  planetary albedos.}

Due to limitations of the heavy-element EoS we did not model very massive metal-rich planets. While our suite of models covers most currently discovered planets (in terms of mass and metallicity), there a few extremely enriched planets \citep[e.g.,][]{Barragan2018} that cannot currently be modelled with \texttt{planetsynth}. An additional limitation of the EoS is that the heavy-elements are represented by a 50-50 rock-ice mixture. However, giant exoplanets are diverse and the heavy elements can differ between planets. Giant planets form at various locations in the protoplanetary disk, which themselves can differ in elemental ratios, and the formation location determines the composition of the accreted solids due to different condensation temperatures. The exact heavy-element composition can affect the planetary radius, by a few percent \citep{Baraffe2008}.

The evolution calculations in this work correspond to the hot-start framework \citep{2003A&A...402..701B,Marley2007}. If planets cool adiabatically, different initial conditions in hot-start models converge quickly \citep{Marley2007,Berardo2017b}. However, this is not true for non-adiabatic contraction, for example, due to the presence of large composition gradients \citep{Vazan2013,Vazan2015} or double-diffusive instabilities \citep{Leconte2012,Wood2013}. While currently it seems that hot-starts are expected in the core accretion framework \citep{Berardo2017a,Berardo2017b,Cumming2018}, future calculations could include cold \citep{Marley2007,2008ApJ...683.1104F} and warm-start initial configurations  \citep{2012ApJ...745..174S,2014MNRAS.437.1378M}. 

Recent studies of the solar system gas giants, Jupiter and Saturn, suggest that their interiors are  not fully convective/adiabatic  \citep[e.g.,][]{Debras2019,2021arXiv210413385M}. Giant planets that are not fully convective cool down more slowly, because radiative/conductive or double-diffusive energy transport is less efficient. As a result, they would have lower luminosities early in their lifetimes because of the less efficient cooling. At some point, however, their luminosities would be higher, because their interiors stay hot for longer. Additional models should therefore ideally also consider non-adiabatic interiors.

\section{Conclusions}\label{sec:conclusions}

In this work we calculate a comprehensive suite of giant planet evolution models with MESA for a large variety of planetary masses, bulk \& atmospheric metallicities and incident stellar fluxes. We present the freely available python module \texttt{planetsynth} (\url{https://github.com/tiny-hippo/planetsynth}) that interpolates on the model grid to generate synthetic evolution tracks. These cooling tracks converge well to those calculated by MESA, with accuracies generally well beyond observational and theoretical uncertainties. 

These synthetic cooling tracks have many scientific applications, including:

\begin{itemize}
    \item Time-dependent mass-radius can easily be generated, which can aid the characterisation of giant planets.
    \item By using mass-radius measurements of exoplanets, cooling tracks can be used to infer a planet's bulk composition.
    \item We show that if observational uncertainties on mass, radius and age diminish further, measurements of the atmospheric composition can be used to constrain the bulk composition.
    \item We demonstrate how the the bulk metallicity and mass of the young directly imaged planet 51 Eri b can be inferred from synthetic cooling tracks.
\end{itemize}

Many accurate and new measurements of giant exoplanets are expected in the upcoming years. For example, JWST will detect young planets in the mid-infrared, which will allow it to image planets with smaller masses than currently possible. Another great advancement will come with the observation of many giant planet atmospheres - in particular their compositions - with the Ariel mission. Combined with the improved determination of stellar ages from Plato and improved measurements from ground-based facilities such as HARPS, NIRPS and ESPRESSO, this will further constrain the compositions of giant exoplanets, opening up compelling avenues to improve our understanding of how giant planets form and evolve. To make of the most of these exciting opportunities, it is essential to consider giant planets as evolving objects. Giant planet evolution models will therefore play a crucial role, and it will be important that these models continue to be refined. The framework for creating synthetic evolution tracks presented in this work is flexible, and it will be possible to include a new generation of giant planet evolution models that account for non-adiabatic cooling histories, composition gradients and other physical mechanisms that can affect the thermal evolution of giant exoplanets.

\section*{Acknowledgements}
We thank Daniel Thorngren for valuable comments that helped to improve the manuscript.
We also acknowledge support from SNSF grant \texttt{\detokenize{200020_188460}} and the National Centre for Competence in Research ‘PlanetS’ supported by SNSF.

\section*{Software}
MESA \citep{Paxton2011,Paxton2013,Paxton2015,Paxton2018,Paxton2019},
NumPy \citep{harris2020array},
SciPy \citep{2020SciPy-NMeth},
Matplotlib \citep{Hunter2007},
scikit-learn \citep{scikit-learn},
Jupyter \citep{jupyter}

\section*{Data Availability}
The data of the evolution models will be shared on request. The python module \texttt{planetsynth} is available at \url{https://github.com/tiny-hippo/planetsynth}.

% \clearpage
\bibliographystyle{mnras}
\bibliography{library}

\appendix
\section{Model Assumptions \& Evolution Calculations}\label{sec:model_assumptions}

\subsection{Creating Planetary Evolution Models with MESA}\label{sec:appendix_model_creation}

The evolution models are calculated in three steps:

\begin{enumerate}
    \item First, we create initial models for the range of planetary masses that we consider using the \texttt{create\_initial\_model} namelist option. We use similar initial radii for all masses, except that we slightly scale them upwards with increasing mass. In this step, we use the same composition for all the planets, except when $Z = 0$. The models are then evolved briefly for a few thousands years for numerical stability.
    \item As the next step, we use the  \texttt{relax\_initial\_composition} namelist option to create the core-envelope profile for the assigned bulk and atmospheric metallicity. This sets the size of the core as well as the metallicity in the envelope and atmosphere. This step was skipped if $Z = 0$. Again, the models are evolved for $10^3$ years.
    \item Lastly, the relaxed models are evolved for $10^{10}$ years with the stellar irradiation added via the \texttt{column\_depth\_for\_irradiation} and \texttt{irradiation\_flux} options. Since we consider stable core-envelope or homogeneous structures, mixing of elements is disabled by setting \texttt{mix\_factor = 0}, and convective regions are determined with the Schwarzschild criterion (\texttt{use\_Ledoux\_criterion = .false.}).
\end{enumerate}

\subsection{Equation of State}\label{sec:appendix_equation_of_state}

Depending on the planetary mass, we use two different versions of MESA. For lower-mass planets ($M \leq 5 M_J$), we use release 10108 with a modified EoS suitable for a wide range of planetary conditions and compositions \citep{Mueller2020}. Our EoS uses the recent \citet{Chabrier2019} hydrogen-helium EoS (hereafter CMS) and combines it with QEoS \citep{More1988,Vazan2013} for rock (SiO$_2$) or water to create ideal mixture of hydrogen, helium and a heavy element. For this work, we implement an additional EoS that uses a 50-50 rock-water mixture, calculated as an ideal mixture. Our heavy-element EoS is valid for densities up to $\rho = 100$ [g/cc], which is not enough for massive giant planets. Therefore, for high-mass planets ($M > 5 M_J$) we use the MESA version 15140 with the built-in MESA EoS, which has the option to use the CMS EoS for H-He. For the conditions relevant in this work, the MESA EoS can be used for metallicities up to $Z = 0.04$, which sets the upper limit of metallicities that we consider for these massive planets.

\subsection{Atmosphere}\label{sec:appendix_atmosphere}

The metallicity in the outer envelope can have a large effect on the evolution via the opacity. This effect is strongest for highly irradiated planets: higher atmospheric metallicities slow the cooling, and yield larger radii and luminosities at a given time \citep{Burrows2007,2020ApJ...903..147M}. Therefore, the envelope metallicity is an important free parameter, which ranges between $Z_{env} = 0$ up to min$(0.1, Z)$ in this work. For the atmospheric model, we use the low-temperature opacities of \citet{Freedman2014} with the \texttt{simple\_photosphere} MESA boundary condition. This defines the photosphere at an optical depth of $\tau = 2/3$.

The irradiated surfaces of the planets are treated with the $F_*-\Sigma_*$ method: the stellar irradiation $F_*$ is applied to a mass column $\Sigma < \Sigma_*$, where $\Sigma(r) = \int_{r}^{R} \rho(r^{\prime}) dr^{\prime}$. \rev{The treatment of the stellar irradiation used in these models can be interpreted in the following way: Irradiation penetrates into the outer layers of the planet, which increases the transported luminosity at the surface.}
Excluding highly irradiated hot Jupiters, this method provides good agreement with semi-analytical models for irradiated planetary atmospheres \citep{Guillot2010,Guillot2011}. We use $\Sigma_* = 3 \times 10^2$ g cm$^{-2}$ in order to match these models, however the planetary evolution is not sensitive to this choice \citep{2015ApJ...813..101V}.

\subsection{Initial Conditions}\label{sec:appendix_initial_conditions}

Calculating the evolution of planets requires suitable initial conditions. We assume that the planets form by core accretion \citep{Mizuno1980,Pollack1996}. In this framework a key assumption is the initial primordial thermal state, which defines the subsequent evolutionary path. Here, we consider so-called \textit{hot-start} models \citep[e.g.,][]{2003A&A...402..701B,Marley2007}. In a hot-start, the initially adiabatic planet is created with an arbitrarily large radius. Alternative core accretion formation models are \textit{cold}- \citep{Marley2007,2008ApJ...683.1104F} or \textit{warm}-start models \citep{2012ApJ...745..174S,2014MNRAS.437.1378M}. For very young planets, predictions from the different models can vary greatly (up to several orders of magnitude in luminosity) \citep[see e.g.,][]{2016PASP..128j2001B}. Therefore, we focus on planets older than $10^7$ years in this work.

\section{Validation Sample \& Error Distributions}\label{sec:appendix_performance_testing}

\begin{figure}
    \centering
    \includegraphics[width=\columnwidth]{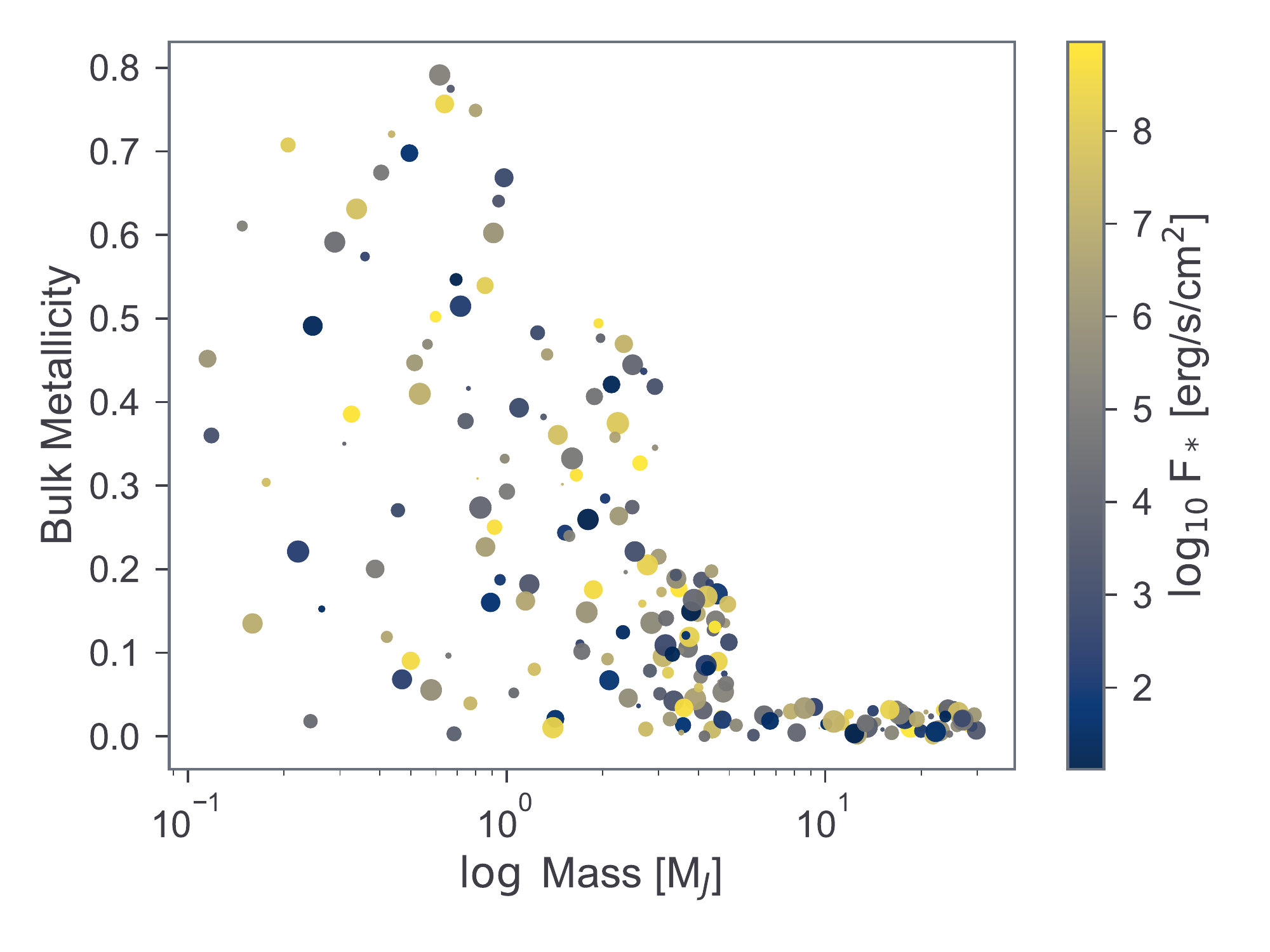}
    \caption{Validation sample (N = 200) generated from the four dimensional parameter space with the Latin Hypercube sampling method. The sizes of the dots correspond to the atmospheric metallicity and the colour shows the planet's incident stellar irradiation.}
    \label{fig:validation_sample}
\end{figure}

In \S \ref{sec:validation} we demonstrate that the synthetic cooling tracks are a good approximation of the actual evolution calculations. In \cref{fig:validation_sample} we show the validation sample in the $\log M - Z$ space; the size and colour of the scatter dots depicting their envelope metallicity and incident stellar irradiation, respectively.

\begin{figure}
    \centering
    \includegraphics[width=\columnwidth]{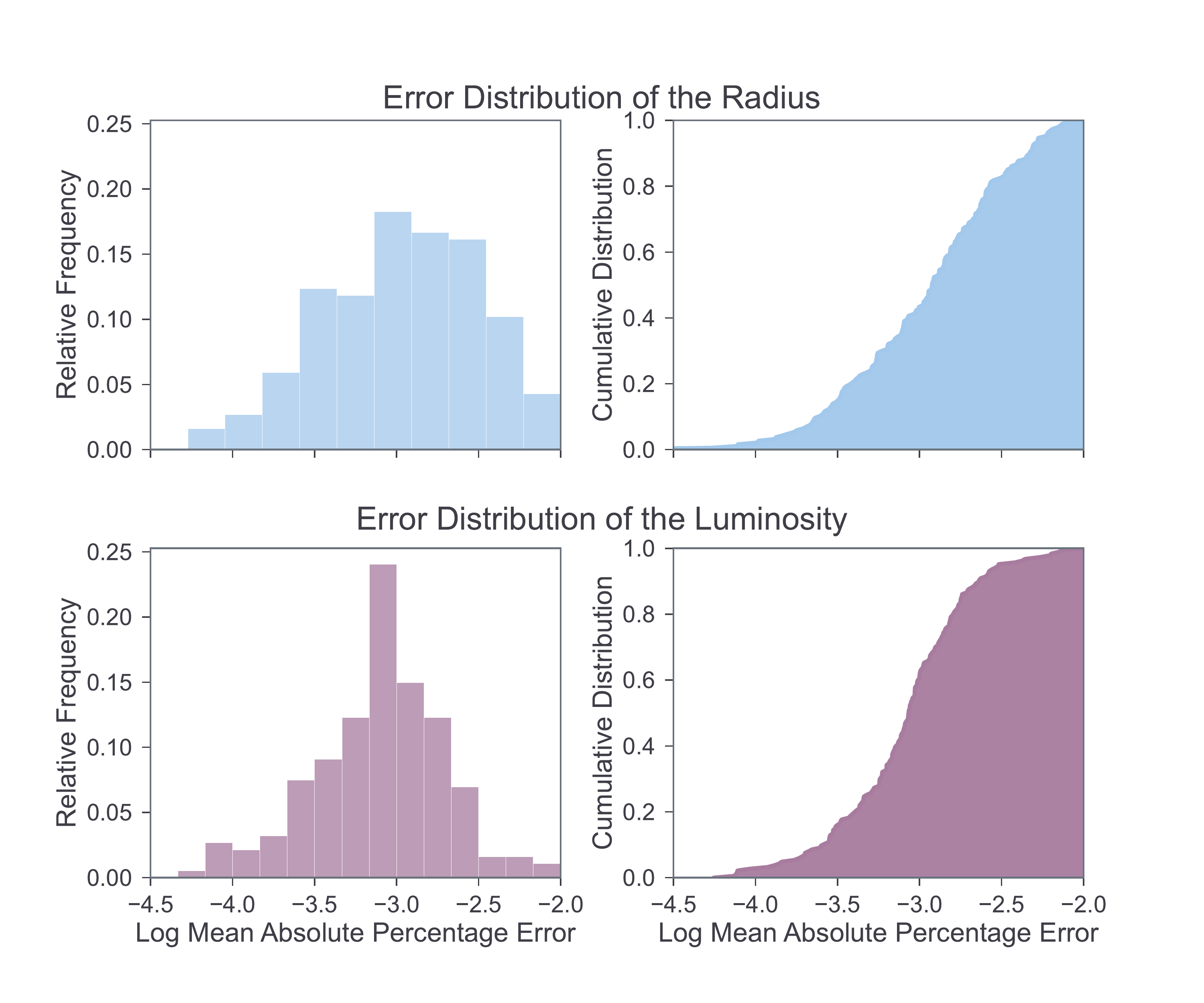}
    \caption{Histograms and cumulative distributions for the $\log_{10}$ MAPE of the radius (top), luminosity (bottom).}
    \label{fig:error_distributions}
\end{figure}

We next present the distributions of mean absolute percentage errors (MAPE) for the radius and luminosity in \cref{fig:error_distributions}. The distributions are approximately Gaussian, slightly skewed to the right. The cumulative distributions show that for almost all the validation samples the MAPE is less than $0.01$\%. As visible in \cref{fig:R_logL_grid}, there are some rare cases where the errors are significant, since they start to become comparable to observational uncertainties. This may occur if the planetary parameters are at or near to the edge of the simulation grid (see Table \ref{table:planet_params}). When interpreting the errors, it is important to keep in mind that there is always an inherent uncertainty in the evolution models due to the assumptions that have to be made.

% \begin{figure*}
%     \centering
%     \includegraphics[width=0.75\textwidth]{img/validation_sample.pdf}
%     \caption{Validation sample (N = 200) generated from the four dimensional parameter space with the Latin Hypercube sampling method. The sizes of the dots correspond to the atmospheric metallicity and the colour shows the planet's incident stellar irradiation.}
%     \label{fig:validation_sample}
% \end{figure*}

% \begin{figure*}
%     \centering
%     \includegraphics[width=0.75\textwidth]{img/error_distributions_mape.pdf}
%     \caption{Histograms and cumulative distributions for the $\log_{10}$ MAPE of the radius (top), luminosity (bottom).}
%     \label{fig:error_distributions}
% \end{figure*}

\bsp	% typesetting comment
\label{lastpage}
\end{document}